# Origin of the Spin-Orbit Interaction

Gianfranco Spavieri[†] and Masud Mansuripur[‡]

[†]*Centro de Física Fundamental, Facultad de Ciencias, Universidad de Los Andes, Mérida, 5101-Venezuela*[*]

[‡]*College of Optical Sciences, The University of Arizona, Tucson, Arizona 85721, USA*

**Abstract**. We consider a semi-classical model to describe the origin of the spin-orbit interaction in a simple system such as the hydrogen atom. The interaction energy $U$ is calculated in the rest-frame of the nucleus, around which an electron, having linear velocity $\boldsymbol{v}$ and magnetic dipole-moment $\boldsymbol{\mu}$, travels in a circular orbit. The interaction energy $U$ is due to the coupling of the induced electric dipole $\boldsymbol{p} = (\boldsymbol{v}/c) \times \boldsymbol{\mu}$ with the electric field $\boldsymbol{E}_n$ of the nucleus. Assuming the radius of the electron's orbit remains constant during a spin-flip transition, our model predicts that the energy of the system changes by $\Delta \mathcal{E} = \frac{1}{2} U$, the factor ½ emerging naturally as a consequence of equilibrium and the change of the kinetic energy of the electron. The correct ½ factor for the spin-orbit coupling energy is thus derived *without* the need to invoke the well-known Thomas precession in the rest-frame of the electron.

**1. Introduction**. The equation for the energy splitting $\Delta \mathcal{E}$ due to spin-orbit interaction was first derived in 1926 by Llewellyn Thomas, using Bohr's model of the hydrogen atom, Schrödinger's quantum mechanics, and relativistic kinematics [1,2]. This result turned out to be in complete agreement with the predictions of Dirac's relativistic quantum mechanics, which was formulated two years later (1928). The Thomas result [3] may be written as

$$\Delta \mathcal{E} = \frac{g}{4m^2 c^2} \frac{dV(r)}{r \, dr} \boldsymbol{S} \cdot \boldsymbol{L}. \tag{1}$$

Here $V(r)$ is the potential energy of the electron at distance $r$ from the nucleus, and $\boldsymbol{L}$ is the orbital angular momentum of the electron, which, in Bohr's classical model, moves in a circular orbit of radius $r$ with velocity $\boldsymbol{v}$ in the presence of the electric field $\boldsymbol{E}_n$ of the nucleus. In the Gaussian system of units, the relation between the magnetic dipole moment $\boldsymbol{\mu}$ and the spin angular momentum $\boldsymbol{s}$ of the electron is

$$\boldsymbol{\mu} = \frac{ge}{2mc} \boldsymbol{s}. \tag{2}$$

In the above equation, $m$ is the mass and $e$ (a negative entity) is the charge of the electron, $c$ is the speed of light in vacuum, and $g \cong 2$ is the $g$-factor associated with the electron's spin magnetic moment. According to Thomas, the interaction energy $U$ between the magnetic moment $\boldsymbol{\mu}$ of the electron and the effective magnetic field $\boldsymbol{B}' \cong -(\boldsymbol{v}/c) \times \boldsymbol{E}_n$ (obtained by a relativistic transformation of the field $\boldsymbol{E}_n$ of the nucleus to the rest-frame of the electron) is

$$U = -\boldsymbol{\mu} \cdot \boldsymbol{B}'. \tag{3}$$

The literature [3,4] describes how, after quantization of $\boldsymbol{s}$ and $\boldsymbol{L}$, and aside from the so-called Thomas factor, the interaction energy $U$ of Eq.(3) assumes the form of $\Delta \mathcal{E}$ given by Eq.(1). For the convenience of the reader, we will derive this result in the following sections.

---

[*]E-mail: spavieri@ula.ve

At the time Thomas proposed his model, it was expected that $U$ had to be added to the basic quantized energy value $\mathcal{E}_n$ obtained via Bohr's model and Schrödinger's wave equation. (Caution: The subscript $n$ of $\mathcal{E}_n$ refers to the $n^\text{th}$ energy level, whereas the subscript $n$ of $\boldsymbol{E}_n$ is a reminder that the $E$-field is that of the nucleus.) Thus, after quantization, $U$ in principle should coincide precisely with the experimentally observed shift in energy given by Eq.(1). The choice of $U$ given by Eq.(3), however, leads to a spin-orbit coupling energy that is twice as large as that in Eq.(1). In order to obtain the correct ½ factor, Thomas resorted to special relativity and tracked the successive relativistic transformations of the electron's rest-frame in its circular orbit. (This relativistic effect is nowadays referred to as the Thomas precession; see Appendix A for a brief derivation of the Thomas precession rate $\Omega_T$ following the elegant approach suggested by E.M. Purcell.) By incorporating the contribution of this relativistic precession, Thomas was able to derive the correct interaction energy, ½$U$, in the rest-frame of the electron.

Although the Thomas ½ factor is fully explained by Dirac's equation, most physics textbooks prefer to describe the spin-orbit interaction in simpler terms using the Thomas model, thus avoiding the more complex formalism of relativistic quantum mechanics. In the words of W.H. Furry [5], "*the original method that was used before the invention of the present quantum mechanics…is lacking in rigor, but it does provide a physical picture for the effect. As long as physics is unfinished business, and physicists must invent approximate models to try to account for unexplained phenomena, the study of arguments of this sort will be important in the physicist's education.*"

Over the years, several authors have attempted to simplify Thomas's argument and presented alternative routes to arriving at the desired ½ factor [5-21]. Some even questioned the validity of Thomas's reasoning (though not that of his final result). In particular, G.P. Fisher [8] states that "*Thomas's original paper…got the correct answer by an incorrect physical argument. This wrong argument persists to this day, so let us hasten to correct it.*" Elsewhere in the same paper [8], Fisher writes: "*Apparently the success of the Dirac equation…made people less interested in probing the details of the atomic spin-orbit interaction. Today one is told that the Thomas effect is included in the Dirac equation. How do we know? Is this another accident*[?]" More recently, Kholmetskii, Missevich, and Yarman [17] have pointed out a "logical inconsistency" in the semi-classical model of spin-orbit splitting, which relies on a non-relativistic equation of motion while including the relativistic phenomenon of Thomas's precession.

In the context of the above history of developments, the present paper aims to introduce an alternative approach to calculating $\Delta\mathcal{E}$ based on a semi-classical model of hydrogen-like atoms (similar to that of Bohr), but with the magnetic dipole-moment $\boldsymbol{\mu}$ of the electron and its related interaction energy $U$ explicitly taken into account. Keeping the treatment in the rest-frame of the nucleus, we show that, in our approach, the factor ½ emerges naturally and without the need to introduce the Thomas precession. Our model represents an alternative to that of Thomas, which, while corroborating his result, provides a simple yet intuitive interpretation of the origin of the spin-orbit interaction energy. The present paper may thus be regarded as a novel application of classical electrodynamics to quantum physics and its interpretation.

Our treatment of spin-orbit coupling in hydrogen-like atoms may be summarized as follows. In its steady-state of motion, the electron revolves around the nucleus in a circular orbit in the $xy$-plane, with its magnetic dipole-moment $\boldsymbol{\mu}$ aligned either parallel or anti-parallel to the $z$-axis. Thus, in the rest-frame of the nucleus, not only does the electron have a magnetic dipole-moment $\mu\hat{\boldsymbol{z}}$, but also a (relativistically-induced) electric dipole-moment $\boldsymbol{p} = p\hat{\boldsymbol{r}} = (\boldsymbol{v}/c) \times \boldsymbol{\mu}$, pointing radially inward or outward, depending on the sign of $\mu$. The nucleus exerts a Coulomb force on the charge $e$ of the electron, as well as a (much weaker) force on the dipoles $\boldsymbol{\mu}$ and $\boldsymbol{p}$. While the Coulomb force is always attractive, the force of the nuclear $E$-field on the dipole pair could be attractive or repulsive, depending on whether $\boldsymbol{\mu}$ is aligned with or against the $z$-axis. We take the orbital radius of our classical electron circling the nucleus to be fixed by the attractive Coulomb force of the nucleus on the charge $e$ of the electron. The perturbing force arising from the action of the nuclear $E$-field on the dipoles $\boldsymbol{\mu}$ and $\boldsymbol{\wp}$ thus affects only the velocity of the electron in its orbit. The resulting change in the kinetic energy of the electron turns out to be one-half the interaction energy $-\boldsymbol{\wp} \cdot \boldsymbol{E}_n$ between the nuclear $E$-field and the dipoles $\boldsymbol{\mu}$ and $\boldsymbol{\wp}$. (Note that there is no magnetic field $\boldsymbol{B}$ in the rest-frame of the nucleus and that, therefore, the interaction energy $-\boldsymbol{\mu} \cdot \boldsymbol{B}$ is zero.) The bottom line is that only one-half of the spin-orbit interaction energy will be available for exchange with an absorbed or emitted photon; the remaining half is needed to adjust the electron's kinetic energy of rotation around the nucleus. This simple mechanism provides the conceptual basis for arriving at the Thomas ½ factor in the rest-frame of the nucleus *without* invoking Thomas's precession.

The search for the origin of the Thomas ½ factor in the rest-frame of the nucleus (including an examination of the role of the induced electric dipole moment $\boldsymbol{\wp}$) has a long and distinguished history. References [12-21] as well as those cited therein provide a good starting point for delving into the subject. In this connection, previous efforts have typically aimed at clarifying the spin dynamics in the rest-frame of the nucleus, thus helping to relate the behavior of the magnetic moment $\boldsymbol{\mu}$ of the electron in its own rest-frame to that in the rest-frame of the nucleus. To our knowledge, no previous investigator has dismissed Thomas's original argument in favor of an alternative mechanism acting directly in the rest-frame of the nucleus; a mechanism that would reduce the spin-orbit energy by the desired ½ factor. In contrast, the premise of the present paper is that Thomas's precession, being a kinematic effect in the rest-frame of the electron, *cannot* account for the observed ½ factor. Instead, we propose that the action of the nuclear $E$-field on $\boldsymbol{\mu}$ and $\boldsymbol{\wp}$ (within the rest-frame of the nucleus) produces a change in the kinetic energy of the electron, which suffices to explain the observed spin-orbit coupling energy.

**2. Energy equation and equilibrium**. In our model, the electron orbits in a circular path of radius $r$ with angular velocity $\omega$ [and linear (tangential) speed $v = r\omega$] in the presence of the field $\boldsymbol{E}_n$ of a massive nucleus having charge $q_n$; see Fig.1. Taking into account the kinetic energy $K$, the potential energy $V$, and the interaction energy $U$ between $\boldsymbol{\mu}$ and $\boldsymbol{E}_n$, the total energy of the system may be written as follows:

$$\mathcal{E} = K + V + U = \tfrac{1}{2} m r^2 \omega^2 + \frac{e q_n}{r} + U. \tag{4}$$

The electromagnetic (EM) interaction energy $U$ can be derived in the rest-frame of the nucleus from the energy expression $U = (8\pi)^{-1} \int (E^2 + B^2) d^3x$, knowing that the moving magnetic dipole $\boldsymbol{\mu}$ possesses an electric dipole $\boldsymbol{\wp} = (\boldsymbol{v}/c) \times \boldsymbol{\mu}$ with its associated electric field $\boldsymbol{E}_\wp$. Therefore, $U = (4\pi)^{-1} \int \boldsymbol{E}_\wp \cdot \boldsymbol{E}_n \, d^3x = -\boldsymbol{\wp} \cdot \boldsymbol{E}_n$, which, for $\boldsymbol{\mu} = \mu \hat{\boldsymbol{z}}$, can be written as

$$U = -\boldsymbol{\wp} \cdot \boldsymbol{E}_n = -[(\boldsymbol{v}/c) \times \boldsymbol{\mu}] \cdot \boldsymbol{E}_n = -r\omega\mu E_n/c. \tag{5}$$

The above equation is equivalent to Eq.(3). It must be pointed out that the expression relating $\boldsymbol{\wp}$ to $\boldsymbol{\mu}$, which is accurate for a particle moving at constant velocity $\boldsymbol{v}$, has been extended here to a case in which the direction of $\boldsymbol{v}$ varies with time. For modest accelerations, such as those associated with the orbital motion of the electron in the hydrogen atom, such an approximation is probably justified.

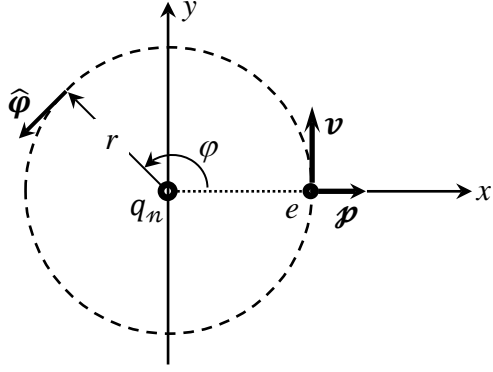

**Fig.1**. An electron orbiting a stationary nucleus of charge $q_n$ located at the origin of coordinates. The orbit's radius is $r$, and the linear velocity of the electron is $\boldsymbol{v} = r\omega\hat{\boldsymbol{\varphi}}$, where $\omega$ is the angular velocity and $\hat{\boldsymbol{\varphi}}$ is the unit-vector in the azimuthal direction. It is assumed that the electron arrives at $(x, y, z) = (r, 0, 0)$ at $t = 0$. The magnetic dipole-moment of the electron (not shown) is $\boldsymbol{\mu} = \mu\hat{\boldsymbol{z}}$, which is perpendicular to the $xy$-plane of the orbit. At $t = 0$, the relativistically-induced electric dipole-moment $\boldsymbol{p} = (\boldsymbol{v}/c) \times \boldsymbol{\mu}$ is aligned with the $x$-axis, as indicated.

**2.1. Force on a moving magnetic dipole and the equilibrium condition**. The basic equilibrium condition requires that the attractive electric force on the electron due to the field $\boldsymbol{E}_n$ of the nucleus, namely, $e\boldsymbol{E}_n = eq_n\hat{\boldsymbol{r}}/r^2$, provide the centripetal acceleration $-(r\omega^2)\hat{\boldsymbol{r}}$. However, there is an additional radial force $\boldsymbol{F}$ acting on $\boldsymbol{\mu}$ which needs to be taken into account. According to the literature [22-25], the expression for $\boldsymbol{F}$ is

$$\boldsymbol{F} = -(\boldsymbol{v}/c) \times (\boldsymbol{\mu} \cdot \boldsymbol{\nabla})\boldsymbol{E}_n. \tag{6}$$

The force $\boldsymbol{F}$ in Eq.(6) corresponds to the motion of a constant $\boldsymbol{\mu}$ in the presence of the external field $\boldsymbol{E}_n$. (This force may be derived by assuming that the electron possesses the electric dipole-moment $\boldsymbol{p}$ and a hidden momentum, to be described in Sec.3. An alternative route to arriving at the same force *without* invoking hidden entities is presented in Sec.4.) In our system, $\boldsymbol{\mu} = \mu\hat{\boldsymbol{z}}$ and, in the $xy$-plane of the orbit, $\partial_z E_x = \partial_z E_y = 0$ while $\partial_z E_z = \partial_z(q_n z/r^3) = q_n/r^3$. Consequently,

$$\boldsymbol{F} = -\left(\frac{r\omega\mu q_n}{cr^3}\right)\hat{\boldsymbol{r}} = -\omega\mu E_n/c = U\hat{\boldsymbol{r}}/r. \tag{7}$$

The equilibrium condition is then,

$$\frac{eq_n}{r^2} + \frac{U}{r} = -mr\omega^2. \tag{8}$$

With the above equilibrium condition established, solving for $\omega$ and substituting it in Eq.(4) yields the energy of our classical system as

$$\mathcal{E} = \frac{eq_n}{2r} + \tfrac{1}{2}U, \tag{9}$$

where the resulting added term $\delta\mathcal{E} = \tfrac{1}{2}U$ emerges naturally with the correct ½ factor included. The interesting physical consequence of Eq.(9) is that if, in a pure spin-flip transition, the orbit radius $r$ remains constant, then the energy of the system will change by $2(\mathcal{E} - \mathcal{E}_n) \cong U$, *without* the need to introduce Thomas's precession mechanism.

Returning now to Eq.(8), let us denote by $r_0$ and $\omega_0$ the orbital radius and the angular velocity of the electron in the absence of spin-orbit coupling (i.e., in the standard Bohr model of the hydrogen atom). Then, considering that the inclusion of spin-orbit interaction associated with $\boldsymbol{\mu} = \mu\hat{\boldsymbol{z}}$ requires only a small adjustment to $r$ and $\omega$, say, by $\delta r = r - r_0$ and $\delta\omega = \omega - \omega_0$, we find, in accordance with Eq.(8), that $\delta(mr^3\omega^2) = -\delta(rU)$, that is,

$$3m\omega_0^2\delta r + 2mr_0\omega_0\delta\omega = -U/r_0. \tag{10}$$

The above equation establishes a link between $\delta r$ and $\delta\omega$ on the one hand, and the interaction energy $U$ on the other, when the magnetic moment $\boldsymbol{\mu}$ of the electron is aligned either with the z-axis (spin down, $U < 0$) or against the z-axis (spin up, $U > 0$). In general, we shall assume that, in a spin-flip process, the orbital radius remains constant at the Bohr radius $r_0$, that is, $\delta r = 0$, and proceed to calculate $\delta\omega$ from Eq.(10). Clearly, the angular velocity of the electron in its spin-up state ($\mu < 0$ and $U > 0$) will be less than that in the standard Bohr model, that is $\omega < \omega_0$, whereas in the spin-down state ($\mu > 0$ and $U < 0$), we will have $\omega > \omega_0$.

Not only does the assumption $\delta r = 0$ lead to the correct (i.e., experimentally observed) value of the spin-orbit energy, but it may also be said to conform to the standard quantum mechanical treatment of the hydrogen atom via Schrödinger's equation. In fact, according to Bohr's semi-classical model, the orbital radius $r$ of the electron is quantized. The Schrödinger equation foresees [26] that the electron wave-function is spread out, and that the orbit radius $r$ is more properly represented by the quantum expectation value $\langle r_{n\ell}\rangle = \frac{1}{2}a_0[3n^2 - \ell(\ell+1)]$, which is a function of the quantum numbers $n$ and $\ell$; here $a_0$ is the Bohr radius of the hydrogen atom. In a pure spin-flip transition, the orbital angular momentum $\boldsymbol{L}$ maintains its quantum number, that is, $\ell$ remains constant. Since also the principal quantum number $n$ remains constant, it might be expected that spin-flip is a transition at constant $r = \langle r_{n\ell}\rangle$.

The assumption that the orbit radius $r$ remains constant during a spin-flip transition in a non-relativistic semi-classical model does not necessarily imply that it holds also for the solution of Dirac's equation in relativistic quantum mechanics. In the context of an extended Bohr model, the spin-dependence of the radial part of the Dirac-Coulomb wave-function has been considered by Kholmetskii, Missevich, and Yarman in [17].

Denoting the particle's energy in the absence of spin-orbit coupling by $\mathcal{E}_n = K_0 + V_0 = eq_n/2r_0$, where $r_0 = n^2\hbar^2/(m|e|q_n)$ is the Bohr radius associated with the principal quantum number $n$, the total classical energy of the system may be written as follows:

$$\mathcal{E} = \mathcal{E}_n + \delta\mathcal{E} = \frac{eq_n}{2r_0} + \frac{1}{2}U = -\frac{me^2q_n^2}{2n^2\hbar^2} - \frac{1}{2}[(\boldsymbol{v}/c)\times\boldsymbol{\mu}]\cdot\boldsymbol{E}_n. \tag{11}$$

Whereas the first term on the right-hand-side of Eq.(11) is the unperturbed energy level $\mathcal{E}_n = -me^2q_n^2/(2n^2\hbar^2)$, the second term, treated as a small perturbation, provides the energy split due to spin-orbit interaction, in agreement with observation.

**2.2. Thomas's contemporaries equated the interaction energy $U$ with the spin-orbit energy**. We speculate now as to why, at the time of Thomas, physicists incorrectly assumed that the interaction energy $U$ had to correspond to the spin-orbit energy $\Delta\mathcal{E}$. We start by pointing out that the first term on the right-hand-side of Eq.(9) is precisely the result of Bohr's model [27], according to which an integer number $n$ of deBroglie wavelengths $\lambda = 2\pi\hbar/p_{mech}$ must fit around the electron's circular orbit. ($\boldsymbol{p}_{mech}$ is the mechanical linear momentum of the electron.) This condition leads straightforwardly to quantization of orbital angular momentum, $mr^2\omega = n\hbar$. The other constraint is imposed by the need to obtain the centripetal acceleration from the attractive Coulomb force of the nucleus, i.e., $mr\omega^2 = -eq_n/r^2$. These two independent constraints on the electron's orbit are subsequently solved to yield $r_0 = n^2\hbar^2/(m|e|q_n)$ and $\omega_0 = me^2q_n^2/(n^3\hbar^3)$. Thus, in the absence of spin-orbit interaction, the kinetic plus potential energy of the electron is found to be

$$\mathcal{E}_n = K_0 + V_0 = \frac{eq_n}{2r_0} = -\tfrac{1}{2}m\left(\frac{eq_n}{n\hbar}\right)^2. \tag{12}$$

Bohr's quantization of angular momentum thus leads to quantization of the orbits and of the energy, in (partial) agreement with the predictions of Schrödinger's equation. Thomas and his contemporaries apparently did not realize that the force $\boldsymbol{F}$ of Eq.(6) acts on the electron. Therefore, the equilibrium condition used in conjunction with the Bohr model lacked the $U$-dependent term in our Eq.(8). For a spin-flip transition occurring at fixed $r$, the potential energy $V(r)$ is obviously constant but, without the $U$-dependent term, the equilibrium condition, Eq.(10), implies that $\delta\omega = 0$, so that also $K$ and, consequently, $\mathcal{E}_n = K + V$ must remain constant. In this case, when the interaction term $U$ is added to $K + V$, as in Eq.(4), the energy change appears to be $\delta\mathcal{E} = U$, leading one to believe that the spin-orbit energy $\Delta\mathcal{E}$ must correspond to $\delta\mathcal{E} = U$. Since the interaction energy $U$ is twice as large as the experimentally observed spin-orbit energy split $\Delta\mathcal{E}$, physicists looked for a way to explain the missing ½ factor, as Thomas did with his precession mechanism, which is associated with a continuous rotation of the rest-frame of the electron.

However, in the rest-frame of the nucleus, where we contend that Thomas's precession is absent, the additional radial force $\boldsymbol{F}$ of Eq.(7) needs to be taken into account [17-19]. In fact, if $U$ is added to $K + V$ while keeping $\delta r = 0$, Eq.(8) dictates that $\omega$ (and also $K$) appearing in Eq.(4) must change—if the equilibrium condition for the orbiting electron is to be restored. Consequently, the resulting change in energy, $\delta\mathcal{E}$, is not $U$, but rather ½$U$, as in Eq.(9).

Let us consider, for example, a spin-flip transition in which the magnetic dipole moment along the $z$-axis flips from $-\boldsymbol{\mu}$ to $\boldsymbol{\mu}$, while the orbit radius $r$ remains constant. Although the interaction energy between the nuclear $E$-field and the relativistically-induced electric dipole $\boldsymbol{p}$ decreases by $2|U|$, the corresponding drop in energy in accordance with Eq.(9) is $\delta\mathcal{E} = -|U|$, where $\delta\mathcal{E}$ correctly represents the corresponding spin-orbit energy change $\Delta\mathcal{E}$ of Eq.(1). In this case, for the energy equation, Eq.(4), to hold while the electron's orbit remains stable at a constant $r$, a change in the kinetic energy $K = \tfrac{1}{2}mr^2\omega^2$ would be mandatory. This change in $K$ can be readily evaluated by invoking the condition $\delta r = 0$ and calculating $\delta\omega$ from Eq.(10).

Our model, of course, cannot predict the quantization of $\mathcal{E}$, $\boldsymbol{L}$, and $\boldsymbol{s}$, nor can it explain why the spin-flip transition must occur at a constant $r$. However, a spin-flip transition at constant $r$ mirrors the observed behavior, whereas one involving a change in the orbital radius does not.

**3. Energy and angular momentum**. In this section we merely point out how $\delta\mathcal{E} = \tfrac{1}{2}U$ can be expressed in the form of $\Delta\mathcal{E}$ given by Eq.(1), and also remark on the changes in the angular momentum of the system. During a spin-orbit process (involving spin-flip and/or a change in the orbital angular momentum), the atom interacts with the EM field (i.e., a photon), exchanging energy and angular momentum with it. An analysis of the dynamical process, which requires the knowledge of forces and torques acting on $\boldsymbol{\mu}$ *during* the transition, is beyond the scope of the present article. The overall change in the energy and angular momentum of the system, however, may be determined by examining the steady states of the system before and after the transition.

The interaction energy $U$ between the relativistically-induced electric dipole-moment $\boldsymbol{p}$ and the field $\boldsymbol{E}_n$ of the nucleus given by Eq.(5) may be written in alternative forms, as follows:

$$U = -[(\boldsymbol{v}/c) \times \boldsymbol{\mu}] \cdot \boldsymbol{E}_n = -\boldsymbol{\mu} \cdot (\boldsymbol{E}_n \times \boldsymbol{v}/c) = -\boldsymbol{v} \cdot (\boldsymbol{\mu} \times \boldsymbol{E}_n/c). \tag{13}$$

As an aside, it is interesting to observe from the last expression in the above equation that $U$ can be expressed in terms of the coupling $-\boldsymbol{v} \cdot \boldsymbol{P}_h$, where $\boldsymbol{P}_h = \boldsymbol{\mu} \times \boldsymbol{E}_n/c$ is the so-called "hidden momentum" of the magnetic dipole in the presence of an external electric field [22-25].

To show now that $\delta\mathcal{E} = \tfrac{1}{2}U$ goes over to $\Delta\mathcal{E}$ of Eq.(1), we recall that, in its non-relativistic approximation, Dirac's equation provides the spin-orbit energy in the following form [28]:

$$\Delta\mathcal{E} = -\frac{e\hbar}{4m^2c^2}(2\boldsymbol{s}/\hbar) \cdot (\boldsymbol{E}_n \times \boldsymbol{p}_{mech}). \tag{14}$$

Here $\boldsymbol{p}_{mech} = m\boldsymbol{v}$ is the momentum of the electron, and $\boldsymbol{E}_n \times \boldsymbol{p}_{mech} = q_n \boldsymbol{r} \times \boldsymbol{p}_{mech}/r^3 = q_n \boldsymbol{L}_{mech}/r^3$. Noting that $eq_n/r^3 = -r^{-1}dV(r)/dr$, it is seen that Eq.(14) reduces to Eq.(1) provided that $g = 2$. This spin-dependent term of the Dirac equation related to the kinetic energy arises from the coupling of the Pauli matrices $\hat{\boldsymbol{\sigma}} (= 2\hat{\boldsymbol{s}}/\hbar)$ with the cross-product $\boldsymbol{E}_n \times \hat{\boldsymbol{p}}$.

Choosing the second to last expression in Eq.(13) and taking into account Eq.(2), the spin-orbit energy in accordance with our model will be

$$\delta\mathcal{E} = \tfrac{1}{2}U = -\tfrac{1}{2}\boldsymbol{\mu} \cdot (\boldsymbol{E}_n \times \boldsymbol{v}/c) = -\frac{ge}{4m^2c^2}\boldsymbol{s} \cdot (\boldsymbol{E}_n \times \boldsymbol{p}_{mech}). \tag{15}$$

This is in agreement with Eq.(14), which is equivalent to the standard expression of spin-orbit energy given by Eq.(1). Recalling Eq.(11), we may now write the Schrödinger Hamiltonian for an electron of mass $m$, charge $e$, and gyromagnetic factor $g \cong 2$, orbiting at radius $r$ in the electric field $\boldsymbol{E}_n = -\boldsymbol{\nabla}\Phi$ of the nucleus, where $V(r) = e\Phi(r) = eq_n/r$, as follows:

$$\widehat{H} = \frac{\hat{\boldsymbol{p}} \cdot \hat{\boldsymbol{p}}}{2m} + e\Phi - \frac{e\hbar}{4m^2c^2}\hat{\boldsymbol{\sigma}} \cdot (\boldsymbol{E}_n \times \hat{\boldsymbol{p}}) = \frac{\hat{\boldsymbol{p}} \cdot \hat{\boldsymbol{p}}}{2m} + e\Phi + \frac{g}{4m^2c^2}\frac{dV(r)}{rdr}\hat{\boldsymbol{s}} \cdot \hat{\boldsymbol{L}}. \tag{16}$$

Switching now to the subject of EM linear and angular momenta, it is well known [22-25] that, when the magnetic dipole $\boldsymbol{\mu}$ of the electron interacts with the electric field $\boldsymbol{E}_n$ of the nucleus, the system acquires an EM momentum

$$\boldsymbol{P}_{em} = (4\pi c)^{-1}\int(\boldsymbol{E} \times \boldsymbol{B})dV = -\boldsymbol{\mu} \times \boldsymbol{E}_n/c, \tag{17}$$

as well as the so-called *hidden* momentum

$$\boldsymbol{P}_h = \boldsymbol{\mu} \times \boldsymbol{E}_n/c = -\boldsymbol{P}_{em}. \tag{18}$$

In the standard Lorentz formulation of classical electrodynamics [29], $\boldsymbol{P}_h$ is interpreted as due to the internal stresses of systems with complex dynamical structure—which is the case for the magnetic dipole in the present situation. In contrast, both $\boldsymbol{P}_{em}$ and $\boldsymbol{P}_h$ are absent in the Einstein-Laub formulation [30], where the EM momentum has a different definition. We will discuss the spin-orbit problem from the perspective of the Einstein-Laub formalism in Sec.4. For the moment, however, it suffices to point out that both formulations yield the same results for the force and torque exerted by $\boldsymbol{E}_n$ on the magnetic dipole $\boldsymbol{\mu}$ of the revolving electron.

Due to the EM fields and their associated stresses, our system thus possesses an EM angular momentum, $\boldsymbol{L}_{em}$, and a hidden angular momentum, $\boldsymbol{L}_h = \boldsymbol{r} \times \boldsymbol{P}_h = \boldsymbol{r} \times (\boldsymbol{\mu} \times \boldsymbol{E}_n/c)$, above and beyond its mechanical orbital angular momentum $\boldsymbol{L}_{mech}$ and spin angular momentum $\boldsymbol{s}$. In a spin-flip process, aside from the change $\Delta s_z = \pm\hbar$, the change of the angular momentum $\boldsymbol{L}$ of the system (around the z-axis) is

$$\delta\boldsymbol{L} = \delta\boldsymbol{L}_{mech} + \delta\boldsymbol{L}_{em} + \delta\boldsymbol{L}_h. \tag{19}$$

There is, therefore, a change in $L_{mech} = (mr^2\omega)\hat{z}$ because of the change $\delta\omega$ in the electron's orbital velocity in accordance with Eq.(10), and also a change in $L_{em}$ and $L_h$, because $\mu$ changes orientation. Conservation of angular momentum requires that $\delta L$ be balanced by the exchange of angular momentum with the absorbed/emitted photon. However, as will be shown in the following section, $\delta L_z$ of Eq.(19) turns out to be much smaller than the change $\Delta s_z = \hbar$ which takes place in consequence of the electron's reversal of spin (from $s = -\tfrac{1}{2}\hbar\hat{z}$ to $s = +\tfrac{1}{2}\hbar\hat{z}$, or vice-versa). Therefore, compared to the change in the spin angular momentum of the system, the change in $L$ given by Eq.(19) is expected to be negligible.

**4. Spin-orbit interaction energy derived in the Einstein-Laub formalism.** We mentioned earlier that the Lorentz and Einstein-Laub formulations lead to identical results. It is worthwhile, therefore, to show explicitly that the force $F$ in Eqs.(6) and (7) is formulation-independent. The present section is devoted to an analysis of the spin-orbit interaction using the Einstein-Laub expressions of force-density and torque-density exerted by an external $E$-field on a moving magnetic dipole [30]. Since there are no hidden entities in the Einstein-Laub formalism, the calculations are fairly straightforward. We switch to the *SI* system of units, so that the reader may see the various formulas in both Gaussian (previous sections) and *SI* systems.

With reference to Fig.1, consider an electron having mass $m$, charge $e$ (negative entity), and magnetic dipole-moment $\boldsymbol{\mu} = \mu\hat{z}$, rotating with angular velocity $\omega$ in a circular orbit of radius $r$ around a massive, stationary nucleus of charge $q_n$. The orbit of the electron is in the $xy$-plane, the nucleus is at $(x_n, y_n, z_n) = (0,0,0)$, and the magnetic moment $\boldsymbol{\mu}$ is either along $\hat{z}$ or $-\hat{z}$, indicating that we are interested only in the two orientations of $\boldsymbol{\mu}$ corresponding to spin-up and spin-down states of the revolving electron. In the *SI* system of units used throughout the present section, the permittivity and permeability of free space are denoted by $\varepsilon_0$ and $\mu_0$, respectively. The various entities needed in our analysis will now be described.

i) The electric field of the nucleus at and around the location of the electron is given by Coulomb's law, as follows:

$$\boldsymbol{E}_n(\boldsymbol{r}) = \frac{q_n \hat{r}}{4\pi\varepsilon_0 r^2} = \left(\frac{q_n}{4\pi\varepsilon_0}\right)\frac{x\hat{x} + y\hat{y} + z\hat{z}}{(x^2+y^2+z^2)^{3/2}}. \qquad (20)$$

ii) The relativistically-induced electric dipole-moment due to the motion of the electron (magnetic dipole-moment $\boldsymbol{\mu} = \mu\hat{z}$, linear velocity $\boldsymbol{v} = r\omega\hat{\varphi}$) is given by

$$\boldsymbol{\wp} = (\varepsilon_0 r\omega\mu)\hat{r}. \qquad (21)$$

(As before, it is being assumed here that the time-dependence of the velocity $\boldsymbol{v}$ does not alter the relation connecting $\boldsymbol{\wp}$ to $\boldsymbol{\mu}$ and $\boldsymbol{v}$.) Recall that in *SI*, where $\boldsymbol{B} = \mu_0 \boldsymbol{H} + \boldsymbol{M}$, the relation between the electron's magnetic dipole-moment $\boldsymbol{\mu}$ and its spin angular momentum $\boldsymbol{s}$ is

$$\boldsymbol{\mu} = \mu_0\left(\frac{ge}{2m}\right)\boldsymbol{s}. \qquad (22)$$

iii) Under the circumstances, the electrostatic interaction energy between the nucleus and the induced electric dipole will be

$$U(r,\omega) = -\boldsymbol{\wp}\cdot\boldsymbol{E}_n = -\frac{\omega\mu q_n}{4\pi r}. \qquad (23)$$

iv) Finally, the electron's kinetic energy $K$, potential energy $V$, mechanical (orbital) angular momentum $\boldsymbol{L}_{mech}$, and the centripetal force $\boldsymbol{F}_c$ acting on the electron may be written as

$$K(r,\omega) = \tfrac{1}{2}mr^2\omega^2. \tag{24}$$

$$V(r) = \frac{q_n e}{4\pi\varepsilon_0 r}. \tag{25}$$

$$\boldsymbol{L}_{mech} = (mr^2\omega)\hat{\boldsymbol{z}}. \tag{26}$$

$$\boldsymbol{F}_c = -(mr\omega^2)\hat{\boldsymbol{r}}. \tag{27}$$

In the Einstein-Laub formalism, the force-density $\boldsymbol{F}(\boldsymbol{r},t)$ and the torque-density $\boldsymbol{T}(\boldsymbol{r},t)$ exerted by the EM fields $\boldsymbol{E}(\boldsymbol{r},t)$ and $\boldsymbol{H}(\boldsymbol{r},t)$ on a material medium specified by its free charge-density $\rho_{\text{free}}(\boldsymbol{r},t)$, free current-density $\boldsymbol{J}_{\text{free}}(\boldsymbol{r},t)$, polarization $\boldsymbol{P}(\boldsymbol{r},t)$, and magnetization $\boldsymbol{M}(\boldsymbol{r},t)$, are written [29-31]

$$\boldsymbol{F}(\boldsymbol{r},t) = \rho_{\text{free}}\boldsymbol{E} + \boldsymbol{J}_{\text{free}} \times \mu_0\boldsymbol{H} + (\boldsymbol{P}\cdot\boldsymbol{\nabla})\boldsymbol{E} + \partial_t\boldsymbol{P}\times\mu_0\boldsymbol{H} + (\boldsymbol{M}\cdot\boldsymbol{\nabla})\boldsymbol{H} - \partial_t\boldsymbol{M}\times\varepsilon_0\boldsymbol{E}, \tag{28a}$$

$$\boldsymbol{T}(\boldsymbol{r},t) = \boldsymbol{r}\times\boldsymbol{F} + \boldsymbol{P}\times\boldsymbol{E} + \boldsymbol{M}\times\boldsymbol{H}. \tag{28b}$$

In the present problem, there is no external magnetic field in the rest-frame of the nucleus, that is, $\boldsymbol{H}(\boldsymbol{r},t) = 0$. Also, the external $E$-field acting on the electron is $\boldsymbol{E}(\boldsymbol{r},t) = \boldsymbol{E}_n(\boldsymbol{r})$, given by Eq.(20). Assuming the electron arrives at $(x,y,z) = (r,0,0)$ at $t = 0$, in the immediate vicinity of $t = 0$, the free charge-density, polarization, and magnetization may be written as

$$\rho_{\text{free}}(\boldsymbol{r},t) = e\delta(x-r)\delta(y-vt)\delta(z), \tag{29}$$

$$\boldsymbol{P}(\boldsymbol{r},t) = (\varepsilon_0 r\omega\mu)\delta(x-r)\delta(y-vt)\delta(z)\hat{\boldsymbol{x}}, \tag{30}$$

$$\boldsymbol{M}(\boldsymbol{r},t) = \mu\delta(x-r)\delta(y-vt)\delta(z)\hat{\boldsymbol{z}}. \tag{31}$$

Appendix B shows that, after algebraic manipulations, the Einstein-Laub force exerted by the nucleus on the moving electron turns out to be

$$\boldsymbol{F}_n(t=0) = \iiint_{-\infty}^{\infty}\boldsymbol{F}(\boldsymbol{r},t=0)dxdydz = \left(\frac{eq_n}{4\pi\varepsilon_0 r^2}\right)\hat{\boldsymbol{x}} - \left(\frac{\omega\mu q_n}{4\pi r^2}\right)\hat{\boldsymbol{x}}. \tag{32}$$

The first term on the right-hand-side of Eq.(32) is the attractive Coulomb force of the nucleus acting on the charge $e$ of the electron. The second term corresponds to the force exerted by the nucleus on the moving magnetic dipole $\boldsymbol{\mu} = \mu\hat{\boldsymbol{z}}$ (including the contribution to the force by the relativistically-induced electric dipole $\boldsymbol{p}$). These are precisely the same forces as obtained in Sec.2 using the Lorentz formalism. (General formulas for the electromagnetic force and torque acting on the revolving electron when $\boldsymbol{\mu}$ is *not* aligned with the $z$-axis are given in Appendix C.)

The force of the nucleus on the electric and magnetic dipoles, i.e., the second term on the right-hand-side of Eq.(32), is much smaller than the force on the charge of the electron (i.e., the first term). Therefore, the contributions of $\boldsymbol{p}$ and $\boldsymbol{\mu}$, the electric and magnetic dipole-moments of the electron, to the central force can alter the orbital motion only slightly. Since the central force $\boldsymbol{F}_n$ of Eq.(32) must be equal to the centripetal force $\boldsymbol{F}_c$ of Eq.(27), we have

$$\frac{eq_n}{4\pi\varepsilon_0 r^2} - \frac{\omega\mu q_n}{4\pi r^2} = -mr\omega^2. \tag{33}$$

In light of the above force-balance equation, invoking Eqs.(22-26), and using the identity $\mu_0\varepsilon_0 = 1/c^2$, the total energy of the electron may now be expressed as

$$\mathcal{E} = K + V + U = \tfrac{1}{2}mr^2\omega^2 + \frac{q_n e}{4\pi\varepsilon_0 r} - \frac{\omega\mu q_n}{4\pi r}$$

$$= \frac{q_n e}{8\pi\varepsilon_0 r} - \frac{\omega\mu q_n}{8\pi r}$$

$$= \tfrac{1}{2}V(r) - \frac{1}{2}\frac{\omega q_n}{4\pi r}\left(\frac{\mu_0 g e}{2m}\right) s$$

$$= \tfrac{1}{2}V(r) + \frac{g}{4m^2 c^2}\frac{dV(r)}{r dr}\mathbf{S}\cdot\mathbf{L}. \tag{34}$$

The above equation indicates that if, during a spin-flip, the radius $r$ of the orbit remains constant, the change in energy will be given by the Thomas formula, Eq.(1)—the factor ½ is thus fully accounted for *without* the need to invoke the precession of the electron's rest-frame.

In general, one might argue that both $r$ and $\omega$ could change during a spin-flip. In that case, Eq.(33) indicates that $mr^3\omega^2$, which in the absence of the spin magnetic moment is equal to $-eq_n/(4\pi\varepsilon_0)$, must increase by $\omega\mu q_n/4\pi$ when $\boldsymbol{\mu} = \mu\hat{\mathbf{z}}$ is aligned with the z-axis (i.e., $\mu > 0$). Alternatively, when $\mu < 0$ (i.e., $\boldsymbol{\mu}$ aligned with the negative z-axis), the quantity $mr^3\omega^2$ must decrease by $\omega\mu q_n/4\pi$. Thus, if the change in the orbital motion of the electron is brought about by a concurrent change in $r$ and $\omega$ (by the small amounts $\delta r = r - r_0$ and $\delta\omega = \omega - \omega_0$), we must have

$$\delta(mr^3\omega^2) = 3mr_0^2\omega_0^2\delta r + 2mr_0^3\omega_0\delta\omega = \frac{\omega_0\mu q_n}{4\pi} \quad \to \quad 1.5\omega_0\delta r + r_0\delta\omega = \frac{q_n\mu}{8\pi mr_0^2}. \tag{35}$$

Any choices for $\delta r$ and $\delta\omega$ that satisfy Eq.(35) would then be acceptable; however, unless $\delta r = 0$, the corresponding change in energy, $\delta\mathcal{E}$, will *not* coincide with the experimentally observed spin-orbit energy given by Eq.(1).

A final remark about the angular momentum of the system is in order. In a spin-flip process, the orbital angular momentum $\mathbf{L}_{mech}$ of the electron will change in accordance with the formula

$$\delta\mathbf{L}_{mech} = \delta(mr^2\omega)\hat{\mathbf{z}} = (2mr_0\omega_0\delta r + mr_0^2\delta\omega)\hat{\mathbf{z}}. \tag{36}$$

Assuming $\delta r = 0$ and using $\delta\omega = q_n\mu/(8\pi mr_0^3)$ from Eq.(35), we find $\delta\mathbf{L}_{mech} = q_n\mu\,\hat{\mathbf{z}}/(8\pi r_0)$. Invoking Eqs.(22) and (33), it is now easy to show that the change of $L_{z\_mech}$ in consequence of a switch from $-\mu\hat{\mathbf{z}}$ to $+\mu\hat{\mathbf{z}}$ (i.e., spin-up to spin-down transition) is given by

$$\delta L_{z\_mech} = \frac{q_n|\mu|}{4\pi r_0} = \left(\frac{g}{2}\right)\left(\frac{v}{c}\right)^2 |s|. \tag{37}$$

Considering that, for the revolving electron, $g \cong 2$, $v \ll c$, and $s = \pm\hbar/2$, it is seen that the change of the mechanical angular momentum in the spin-flip process is much smaller than $\hbar$.

In addition to the mechanical angular momentum just mentioned, the spin-flip process is accompanied by a change in the EM angular momentum of the system. In the Einstein-Laub formalism, the EM momentum-density is $\boldsymbol{p}_{em}(\boldsymbol{r},t) = \boldsymbol{E}(\boldsymbol{r},t) \times \boldsymbol{H}(\boldsymbol{r},t)/c^2$ [29]. For a point-charge–point-magnet system such as a stationary nucleus of charge $q_n$ at a distance $r$ from a stationary electron whose magnetic moment is $\mu\hat{\mathbf{z}}$, it is not difficult to show that [32]

$$\iiint_{-\infty}^{\infty} \boldsymbol{p}_{em}(\boldsymbol{r})dxdydz = 0. \tag{38}$$

Consequently, no net EM momentum resides in the system. However, the system's total EM angular momentum does *not* vanish. One can show that [32]

$$\boldsymbol{L}_{em} = \iiint_{-\infty}^{\infty} \boldsymbol{r}\times\boldsymbol{p}_{em}(\boldsymbol{r})dxdydz = (\boldsymbol{r}_e - \boldsymbol{r}_n)\times(\boldsymbol{\mu}\times\varepsilon_0\boldsymbol{E}_n) = \left(\frac{q_n\mu}{4\pi r}\right)\hat{\mathbf{z}}. \tag{39}$$

A transition from spin-up to spin-down thus raises the EM angular momentum of the system by $2q_n|\mu|\hat{\mathbf{z}}/(4\pi r_0)$, which is twice as large as the corresponding change in the mechanical

angular momentum given by Eq.(37). Once again, such changes in the angular momentum of the system are negligible compared to the change in the spin angular momentum $\Delta \boldsymbol{s} = \pm \hbar \hat{\boldsymbol{z}}$.

**5. Concluding remarks**. We have considered a model in which the spin-orbit energy originates in the overlap of the electric field $\boldsymbol{E}_{\wp}$ of the relativistically-induced electric dipole $\boldsymbol{\wp}$ with the field $\boldsymbol{E}_n$ of the nucleus, yielding, in the rest-frame of the nucleus, the interaction energy $U = -\boldsymbol{\wp} \cdot \boldsymbol{E}_n$ of Eq.(5). Assuming that in a spin-flip transition the radius $r$ of the orbit remains constant, our model predicts that, upon introducing the interaction energy $U$, the overall energy of the system changes by $\delta \mathcal{E} = \frac{1}{2} U$, as appears in Eq.(9). The crucial ½ factor originates from the concurrent change in the kinetic energy $K$ of the revolving electron during the spin-flip process, emerging naturally as a consequence of the effect of the radial force $\boldsymbol{F}$ on the stability of the orbit; see Eq.(7). In the end, only ½ of the interaction energy $U$ is available to be exchanged with the absorbed or emitted photon, that is, $\delta \mathcal{E} = \frac{1}{2} U$ goes over to $\Delta \mathcal{E}$ of Eq.(1).

By failing to incorporate the radial force $\boldsymbol{F} = U \hat{\boldsymbol{r}}/r$ of Eq.(7) into the equilibrium condition, Thomas's contemporaries were led to incorrectly believe that $\Delta \mathcal{E}$ for the Bohr model of the hydrogen atom (and the corresponding solution to Schrödinger's equation) had to be equated with the additional interaction energy $U$ given by Eq.(3). The neglect of $\boldsymbol{F}$ thus resulted in a theoretical over-estimation of the expected spin-orbit energy by a factor of 2, which was subsequently claimed to be corrected by L. Thomas [1,2]. The present paper has argued that the correct ½ factor may be derived *without* resort to Thomas's precession, requiring only that the kinetic energy $K$, the potential energy $V$, and the interaction energy $U$ between the electron's magnetic moment $\boldsymbol{\mu}$ and the nuclear $E$-field be calculated in the rest-frame of the nucleus—with the caveat that, in a spin-flip transition, the orbit of the electron must maintain a constant radius.

The question asked by G.P. Fisher [8] (and mentioned in our introductory section) remains as to whether the agreement between Thomas's result and relativistic quantum mechanics is an accident. Either way, the spin-orbit energy calculated in the electron's rest-frame must be corroborated with the corresponding calculations in the rest-frame of the nucleus. The results of the present paper indicate that Thomas's conclusion (if not his methodology) could be brought into alignment with the spin-orbit energy obtained in the rest-frame of the nucleus. One thing that Thomas's method does not clarify is the fate of the remaining energy, $\frac{1}{2} U = -\frac{1}{2} \boldsymbol{\wp} \cdot \boldsymbol{E}_n$, which is *not* carried by the absorbed/emitted photon. The results of the preceding sections show that the remaining energy goes into (or comes out of) the kinetic energy $K$ of the orbiting electron as seen in the rest-frame of the nucleus.

In conclusion, Bohr's model of the hydrogen atom can be extended to account for the observed spin-orbit interaction with the stipulation that, during a spin-flip transition, the orbital radius $r$ remains constant. In other words, if there is a desire to extend Bohr's model to accommodate the spin of the electron, then experimental observations mandate robust orbits during spin-flip transitions. This is tantamount to admitting that Bohr's model is of limited value, and that one should really rely on Dirac's equation for the physical meaning of spin, for the mechanism that gives rise to $g = 2$, for Zeeman splitting, for relativistic corrections to Schrödinger's equation, for Darwin's term, and for the correct ½ factor in the spin-orbit coupling energy. Bohr's model is a poor man's way of understanding the hydrogen atom. If one desires to extend Bohr's model to account for the spin-orbit interaction, then one must introduce the ad hoc assumption that the orbit radius $r$ is invariant during a spin-flip transition. While a strong physical justification in support of this assumption does not seem to exist, it at least provides a plausibility argument for the observed ½ factor.

# Appendix A

Following a suggestion by E. M. Purcell as reported in G. F. Smoot's Berkeley lecture notes [33], we derive Thomas's precession formula for the magnetic moment of an electron revolving with constant angular velocity $\omega$ in a circular orbit of radius $r$ around a stationary nucleus.

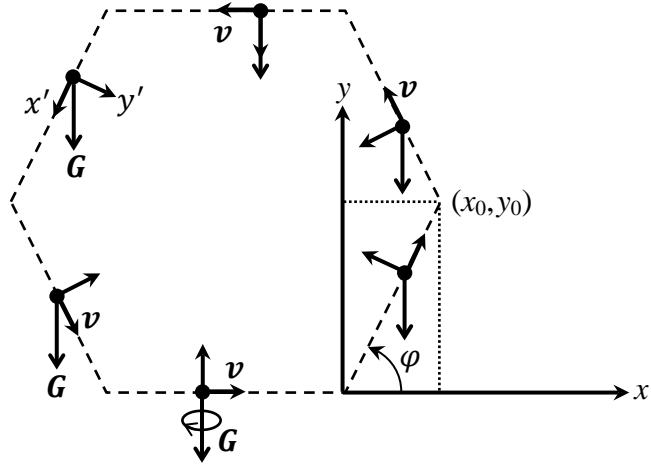

**Fig. A1**. A point-particle, such as an electron, travels with constant speed $v$ around a regular, $n$-sided polygon in the $xy$-plane. Residing in the particle's rest-frame is a gyroscope, whose spin axis $\mathbf{G}$ maintains a constant direction in space as seen in the laboratory frame. At every corner, the particle swiftly changes direction through the angle $\varphi = 2\pi/n$. From the perspective of the particle, however, the turn angle is somewhat greater than $\varphi$ due to the Lorentz-FitzGerald contraction of lengths along the direction of travel. The gyroscope thus appears to undergo a clockwise rotation when viewed in the rest-frame of the particle.

With reference to Fig. A1, let a point-particle travel at constant speed $v = r\omega$ around a regular $n$-sided polygon in the $xy$-plane. Also residing in the particle's rest-frame is a gyroscope, whose spin axis $\mathbf{G}$ maintains a constant orientation within the laboratory frame $xyz$. When, at $t = 0$, the particle arrives at the origin, $(x, y) = (0, 0)$, it suddenly changes direction and moves toward the next vertex located at $(x, y) = (x_0, y_0)$. From the perspective of a stationary observer in the inertial $xyz$ frame, the particle has made a swift turn at an angle $\varphi = \tan^{-1}(y_0/x_0)$. However, in the rest-frame of the particle, at $t' = 0$, the coordinates of the next vertex are $(x_0', y_0') = (\sqrt{1 - v^2/c^2}\, x_0, y_0)$. Therefore, the particle "believes" that it has turned through a somewhat larger angle $\varphi'$, namely,

$$\varphi' = \tan^{-1}\left(\frac{y_0}{\sqrt{1-(v/c)^2}\, x_0}\right). \tag{A1}$$

Suppose now that the polygon has a large number of sides, that is, $n \gg 1$, so that, in the limit of large $n$, it approaches a circle. We may then use the small-angle approximation to write

$$\varphi' \cong \frac{\varphi}{\sqrt{1-(v/c)^2}}. \tag{A2}$$

Now, in the $xyz$ frame, which is the rest-frame of the nucleus, each sharp turn corresponds to $\varphi = 2\pi/n$ radians. However, from the particle's perspective, its direction of travel changes by $\varphi' = 2\pi\gamma/n$ at each turn [using standard relativistic notation $\gamma = 1/\sqrt{1-(v/c)^2}$]. Consequently, when the full circle is traversed, the particle believes that it has turned through a cumulative angle of $n\varphi' = 2\pi\gamma$. This, of course, is an exact result, because, in the limit when $n \to \infty$, the small-angle approximation that led to Eq.(A2) becomes accurate.

In the rest-frame of the electron, the spin axis $G$ of the gyroscope, which is not subject to any external influences, appears to rotate *clockwise* in the $x'y'$-plane; see Fig. A1. Therefore, upon completing one full cycle of revolution around the nucleus, the vector $G$ appears to have undergone a *clockwise* rotation through the angle $\Delta\varphi' = 2\pi(\gamma - 1)$. We may write

$$\Delta\varphi' = 2\pi(\gamma - 1) = 2\pi\left(\frac{\gamma^2-1}{\gamma+1}\right) = 2\pi\left(\frac{\gamma^2}{\gamma+1}\right)\left(1 - \frac{1}{\gamma^2}\right) = 2\pi\left(\frac{\gamma^2}{\gamma+1}\right)\left(\frac{v^2}{c^2}\right). \tag{A3}$$

Recalling that the particle's angular velocity is $\omega = v/r$, the number of full rotations per second around the circle in the $xyz$ frame is $\omega/2\pi$; the same entity in the particle's rest-frame is given by $\gamma\omega/2\pi$ (due to time dilation). Consequently, the apparent precession rate of $G$ around the $z$-axis in the particle's rest-frame is given by

$$\boldsymbol{\Omega}_T = -\omega\left(\frac{\gamma^3}{\gamma+1}\right)\left(\frac{v^2}{c^2}\right)\hat{\boldsymbol{z}} = -\frac{\gamma^3}{\gamma+1}\left(\frac{r^2\omega^3}{c^2}\right)\hat{\boldsymbol{z}} = \frac{\gamma^3}{\gamma+1}\left(\frac{\boldsymbol{a}\times\boldsymbol{v}}{c^2}\right). \tag{A4}$$

In the above expression of the Thomas precession rate, $\boldsymbol{v} = r\omega\hat{\boldsymbol{\varphi}}$ is the linear velocity, and $\boldsymbol{a} = -r\omega^2\hat{\boldsymbol{r}}$ is the radial acceleration of the revolving particle—both measured in the $xyz$ frame. Thomas's precession is thus seen to be a purely geometrical effect rooted in the Lorentz-FitzGerald length-contraction and time-dilation of special relativity. For $v \ll c$, which is typical of atomic hydrogen, we have $\gamma \cong 1$ and $\gamma^3/(\gamma + 1) \cong \tfrac{1}{2}$, leading to

$$\boldsymbol{\Omega}_T \cong -\left(\frac{r^2\omega^3}{2c^2}\right)\hat{\boldsymbol{z}}. \tag{A5}$$

We mention in passing that, in the above discussion, as in much of the literature, the contribution of the Coriolis force to the precession of $G$ (as seen in the particle's rest-frame) has been ignored. This is because in the expression of $\Delta\varphi'$ in Eq.(A3) we discounted the ordinary rotation (per revolution cycle) of the $x'y'$ axes through $2\pi$ radians. The Coriolis force exerts an apparent torque on the gyroscope, which causes a precession of its spin axis at the rate of $\boldsymbol{\Omega}_C = -\omega\hat{\boldsymbol{z}}$. Unlike the relativistic Thomas precession, the non-relativistic precession attributed to the Coriolis force has been deemed incapable of affecting the spin-orbit coupling energy. This is appropriate considering that the Coriolis torque, being fictitious, cannot affect the energy of the gyroscope. However, since the Thomas precession is similar in character to the non-relativistic rotation of coordinates, it is not clear why this relativistic counterpart of the Coriolis torque should be relied upon to arrive at the correction to the spin-orbit coupling energy.

Returning to Thomas's argument, suppose now that $\boldsymbol{\mu}$ represents the magnetic dipole-moment of a particle, being related to its intrinsic angular momentum $\boldsymbol{s}$ via $\boldsymbol{\mu} = (ge/2mc)\boldsymbol{s}$; see Eq.(2). In a constant, uniform magnetic field $\boldsymbol{B}$, the time-rate-of-change of $\boldsymbol{s}$ follows Newton's law, $d\boldsymbol{s}/dt = \boldsymbol{\mu}\times\boldsymbol{B}$, where $\boldsymbol{\mu}\times\boldsymbol{B}$ is the torque exerted by $\boldsymbol{B}$ on the magnetic moment. Since $d\boldsymbol{s}/dt = \boldsymbol{\Omega}\times\boldsymbol{s}$, where $\boldsymbol{\Omega}$ is the precession rate of the dipole-moment around $\boldsymbol{B}$, we find

$$\boldsymbol{\Omega} = -\left(\frac{ge}{2mc}\right)\boldsymbol{B}. \tag{A6}$$

Considering that the dipole's energy in the presence of the $B$-field is $\mathcal{E} = -\boldsymbol{\mu}\cdot\boldsymbol{B} = \boldsymbol{s}\cdot\boldsymbol{\Omega}$, Thomas found it plausible to relate the precession rate $\boldsymbol{\Omega}_T$ of Eq.(A5) to the energy of the revolving electron.

Now, the magnetic field $\boldsymbol{B}' \cong -(\boldsymbol{v}/c)\times\boldsymbol{E}_n$, which appears in Eq.(3) and is produced by a Lorentz transformation of the nuclear field $\boldsymbol{E}_n$ to the rest-frame of the electron, may be written

$$\boldsymbol{B}' \cong -\left(\frac{mr^2\omega^3}{ec}\right)\hat{\boldsymbol{z}}. \tag{A7}$$

To see this, note that $\boldsymbol{v} = r\omega\hat{\boldsymbol{\varphi}}$, $\boldsymbol{E}_n = q_n\hat{\boldsymbol{r}}/r^2$, and $e E_n = -mr\omega^2$. Thus, in accordance with Eq.(A6), the precession frequency associated with $\boldsymbol{B}'$ in the electron's rest-frame is

$$\boldsymbol{\Omega} = -\left(\frac{ge}{2mc}\right)\boldsymbol{B}' \cong \left(\frac{gr^2\omega^3}{2c^2}\right)\hat{\boldsymbol{z}}. \tag{A8}$$

Noting that $g \cong 2$, it is readily seen that the above frequency is twice as large as that associated with the Thomas precession, as given by Eq.(A5); moreover, $\boldsymbol{\Omega}$ and $\boldsymbol{\Omega}_T$ are seen to have opposite signs. That is how Thomas concluded that the energy associated with a spin-flip transition must be one-half of the energy $U$ appearing in Eq.(3), that is, $\Delta\mathcal{E} = \tfrac{1}{2}U = -\tfrac{1}{2}\boldsymbol{\mu} \cdot \boldsymbol{B}'$.

## Appendix B

Using Eqs.(28-31) and with the aid of Eq.(20), we calculate the Einstein-Laub force exerted by the nucleus on the moving electron, as follows:

$$\boldsymbol{F}_n(t=0) = \iiint_{-\infty}^{\infty} \boldsymbol{F}(\boldsymbol{r}, t=0) dxdydz$$

$$= \iiint_{-\infty}^{\infty} [\rho_{\text{free}}\boldsymbol{E} + (\boldsymbol{P} \cdot \boldsymbol{\nabla})\boldsymbol{E} - \partial_t \boldsymbol{M} \times \varepsilon_0 \boldsymbol{E}] dxdydz$$

$$= \iiint_{-\infty}^{\infty} \left\{ \frac{eq_n}{4\pi\varepsilon_0} \delta(x-r)\delta(y-vt)\delta(z) \frac{x\hat{\boldsymbol{x}} + y\hat{\boldsymbol{y}} + z\hat{\boldsymbol{z}}}{(x^2+y^2+z^2)^{3/2}} \right.$$

$$+ \frac{\varepsilon_0 r\omega\mu q_n}{4\pi\varepsilon_0} \delta(x-r)\delta(y-vt)\delta(z) \frac{\partial}{\partial x}\left[\frac{x\hat{\boldsymbol{x}} + y\hat{\boldsymbol{y}} + z\hat{\boldsymbol{z}}}{(x^2+y^2+z^2)^{3/2}}\right]$$

$$\left. + \frac{v\mu q_n}{4\pi} \delta(x-r)\delta'(y-vt)\delta(z)\hat{\boldsymbol{z}} \times \frac{x\hat{\boldsymbol{x}} + y\hat{\boldsymbol{y}} + z\hat{\boldsymbol{z}}}{(x^2+y^2+z^2)^{3/2}} \right\} dxdydz. \tag{B1}$$

Recalling the sifting property of Dirac's delta function and its derivative, namely,

$$\int_{-\infty}^{\infty} f(x)\delta(x-x_0)dx = f(x_0), \tag{B2}$$

$$\int_{-\infty}^{\infty} f(x)\delta'(x-x_0)dx = -f'(x_0), \tag{B3}$$

straightforward algebraic manipulations of Eq.(B1) yield

$$\boldsymbol{F}_n(t=0) = \frac{eq_n\hat{\boldsymbol{x}}}{4\pi\varepsilon_0 r^2}$$

$$+ \frac{r\omega\mu q_n}{4\pi} \iiint_{-\infty}^{\infty} \left[\frac{\hat{\boldsymbol{x}}}{(x^2+y^2+z^2)^{3/2}} - \frac{3x(x\hat{\boldsymbol{x}} + y\hat{\boldsymbol{y}} + z\hat{\boldsymbol{z}})}{(x^2+y^2+z^2)^{5/2}}\right] \delta(x-r)\delta(y)\delta(z) dxdydz$$

$$+ \frac{r\omega\mu q_n}{4\pi} \iiint_{-\infty}^{\infty} \frac{x\hat{\boldsymbol{y}} - y\hat{\boldsymbol{x}}}{(x^2+y^2+z^2)^{3/2}} \delta(x-r)\delta'(y)\delta(z) dxdydz$$

$$= \frac{eq_n\hat{\boldsymbol{x}}}{4\pi\varepsilon_0 r^2} - \frac{2r\omega\mu q_n\hat{\boldsymbol{x}}}{4\pi r^3} + \frac{r\omega\mu q_n}{4\pi} \int_{-\infty}^{\infty} \frac{r\hat{\boldsymbol{y}} - y\hat{\boldsymbol{x}}}{(r^2+y^2)^{3/2}} \delta'(y) dy$$

$$= \frac{eq_n\hat{\boldsymbol{x}}}{4\pi\varepsilon_0 r^2} - \frac{2\omega\mu q_n\hat{\boldsymbol{x}}}{4\pi r^2} + \frac{r\omega\mu q_n}{4\pi} \left[\frac{\hat{\boldsymbol{x}}}{(r^2+y^2)^{3/2}} + \frac{3y(r\hat{\boldsymbol{y}} - y\hat{\boldsymbol{x}})}{(r^2+y^2)^{5/2}}\right]_{y=0}$$

$$= \frac{eq_n\hat{\boldsymbol{x}}}{4\pi\varepsilon_0 r^2} - \frac{2\omega\mu q_n\hat{\boldsymbol{x}}}{4\pi r^2} + \frac{r\omega\mu q_n\hat{\boldsymbol{x}}}{4\pi r^3}$$

$$= \left(\frac{eq_n}{4\pi\varepsilon_0 r^2}\right)\hat{x} - \left(\frac{\omega\mu q_n}{4\pi r^2}\right)\hat{x}. \tag{B4}$$

This is the result that was stated in Eq.(32).

## Appendix C

With reference to Fig.1, if, in the electron's rest-frame, the magnetic moment $\boldsymbol{\mu}$ happens to have components along all three axes, that is, $\boldsymbol{\mu} = \mu_x\hat{x} + \mu_y\hat{y} + \mu_z\hat{z}$, then, in the rest-frame of the nucleus, we will have

$$\boldsymbol{M}(\boldsymbol{r},t) = \left(\mu_x\hat{x} + \gamma^{-1}\mu_y\hat{y} + \mu_z\hat{z}\right)\delta(x-r)\delta(y-vt)\delta(z). \tag{C1}$$

$$\boldsymbol{P}(\boldsymbol{r},t) = \varepsilon_0 v(\mu_z\hat{x} - \mu_x\hat{z})\delta(x-r)\delta(y-vt)\delta(z). \tag{C2}$$

Substitution into Eqs.(28), followed by integration over the volume of the particle, yields

$$\boldsymbol{F}_{EL}(t=0) = \left(\frac{eq_n}{4\pi\varepsilon_0 r^2}\right)\hat{x} - \left(\frac{q_n v}{4\pi r^3}\right)(\mu_z\hat{x} + 2\mu_x\hat{z}). \tag{C3}$$

$$\boldsymbol{T}_{EL}(t=0) = \left(\mu_x\hat{x} + \gamma^{-1}\mu_y\hat{y}\right) \times \left(\frac{q_n v}{4\pi r^2}\right)\hat{z}. \tag{C4}$$

These are the Einstein-Laub force and torque acting on the revolving electron in the rest-frame of the nucleus; the torque is calculated with respect to the instantaneous position of the electron at $t=0$, namely, $\boldsymbol{r}_0 = (r,0,0)$.

In its own rest-frame, the electron is acted upon by $\boldsymbol{F}'_{EL} = \gamma\boldsymbol{F}_{EL}$ and $\boldsymbol{T}'_{EL} = \boldsymbol{\mu} \times \boldsymbol{H}'$, where $\boldsymbol{H}' = (\gamma q_n v/4\pi r^2)\hat{z}$ is the magnetic field of the revolving nucleus. Clearly, $\boldsymbol{T}_{EL} \cong \boldsymbol{T}'_{EL}$ at non-relativistic velocities where $\gamma \cong 1$. This near-equality of $\boldsymbol{T}_{EL}$ and $\boldsymbol{T}'_{EL}$ is perhaps another indication that Thomas's precession mechanism *cannot* be responsible for the ½ factor in the expression of the spin-orbit coupling energy.

**Acknowledgment**. This work has been supported by the CDCHTA, ULA, Mérida, Venezuela.